# Exfoliating pristine black phosphorus down to the monolayer: photo-oxidation and quantum confinement

Alexandre Favron[1], Etienne Gaufrès[2], Fréderic Fossard[3], Pierre L. Lévesque[2], Anne-Laurence Phaneuf-L'Heureux[4], Nathalie Y-W. Tang[2], Annick Loiseau[3], Richard Leonelli[1], Sébastien Francoeur[4] and Richard Martel[2]*

[1] Regroupement Québécois sur les Matériaux de Pointe (RQMP) and Département de physique, Université de Montréal, Montréal QC H3C 3J7, Canada

[2] RQMP and Département de chimie, Université de Montréal, Montréal QC H3C 3J7, Canada

[3] Laboratoire d'Etude des Microstructures, UMR 104 CNRS-Onera, Châtillon, France

[4] RQMP and Département de génie physique, Polytechnique Montréal, Montréal QC H3C 3A7, Canada

*Correspondence to: r.martel@umontreal.ca

**Thin layers of black phosphorus have recently raised interest for their two-dimensional (2D) semiconducting properties, such as tunable direct bandgap and high carrier mobilities. This lamellar crystal of P atoms stacked together by weak van der Waals forces can be exfoliated down to the stratophosphane monolayer (also called phosphorene) using procedures similar to those used for graphene. Properties of this 2D material are however challenging to probe due to a fast and ubiquitous degradation upon exposure to ambient conditions. Herein, we investigate the crystal degradation using *in-situ* Raman and transmission electron spectroscopies and highlight a process involving a photo-induced oxidation reaction with adsorbed oxygen in water. The experimental conditions to prepare and preserve mono-, bi- and multilayers of stratophosphane in their pristine states were determined. Study on these 2D layers provides new insights on the effect of confinement on the chemical reactivity and the vibrational modes of black phosphorus.**

Introduction

Black Phosphorus (P(black)) stands out from the other allotropes of elemental phosphorus, such as the amorphous red phosphorus used to ignite matchsticks by friction and the notoriously unstable white phosphorus polymorph, by its good stability up to 550°C and its anisotropic lamellar structure similar to that of graphite[1,2]. Composed of puckered layers of P atoms stacked together by weak van der Waals forces, P(black) crystals can be exfoliated down to ultrathin layers using techniques similar to those used to prepare graphene from graphite and dichalcogenide monolayers from bulk crystals[3]. Bulk P(black) is a direct bandgap semiconductor ($E_g$=0.35 eV) showing intrinsic electron and hole mobilities at room temperature of 220 and 350 $cm^2V^{-1}s^{-1}$, respectively[4-6]. Few years ago, the properties of exfoliated P(black) have raised interest because of their potential use as anode materials in Lithium-ion batteries[7,8]. In the context of two-dimensional (2D) materials such as graphene and chalcogenides, thickness-dependent carrier mobility in few layers transistors has been reported[9], giving impressive hole mobilities (up to 1000 $cm^2V^{-1} s^{-1}$) and ON/OFF current ratios (up to ~$10^5$). Moreover, a significant blue shift of the bandgap energy with decreasing layer thickness was obtained by *ab-initio* calculations and observed in preliminary luminescence results[10]. Another study combining Raman spectroscopy and atomic force microscopy (AFM) of exfoliated P(black) by mechanical cleavage and plasma treatments has given Raman signatures and optical contrasts that can be used to determine the layer thickness[11], albeit the signature of quantum confinement with the number of layers remains to be further clarified.

It appears then that exfoliated P(black) offers new and exciting prospects for applications in electronics and optoelectronics. However, Bridgman's first synthesis of P(black) dates back as far as 100 years ago and yet the literature on the topic is surprisingly sparse and commercial use of the material is inexistent, which acknowledges major difficulties ahead for its synthesis, processing and manipulation[12,7]. The notoriously poor chemical and structural stability of elemental phosphorus[13] raises then important concerns for all processing steps carried out in ambient conditions. Hence, an important milestone towards the establishment of phosphorus-based electronics is the better understanding of the chemical and physical properties of phosphorus thin layers.

Here, we present a systematic investigation of the chemical reactivity of exfoliated layers of P(black) with oxygen and water in different conditions and provide guidelines on how to prevent oxidation and prepare pristine (oxygen free) materials down to the monolayer. A thorough spectroscopic investigation was first used to determine the origin of the irreversible oxidation of thin layers of P(black) occurring immediately after exfoliation. The experiments revealed a thickness dependent photo-activated charge transfer process with adsorbed oxygen in water under ambient conditions. Control over this ubiquitous photo-oxidation reaction was achieved and provided access to pristine layers for further studies using transmission electron microscopy (TEM), Electron Energy Loss (TEM-EELS) and polarized Raman spectroscopies. TEM images of mono-, bi- and multilayers along with unambiguous signatures in the Raman spectra of electronic confinement effects are shown and discussed using group symmetry analysis.

Because the monolayer is composed of tervalent phosphorus atoms, it belongs, according to the IUPAC nomenclature, to the phosphane group[14,15]. Therefore, a monolayer of exfoliated P(black) is hereafter designated stratophosphane (2D-phosphane), as opposed to the name "phosphorene" used in previous papers, in order to avoid confusion with other chemical derivatives of the materials and other elemental monolayers having $sp^2$- or $sp^3$-hybridization (e.g. graphene, silicene, graphane, germanane, etc.).

Figure 1 presents the main evidences of degradation acquired by AFM and Raman spectroscopy on exfoliated layers produced using a technique derived from the scotch-tape exfoliation technique[16] in order to quantify the morphology and structural changes, respectively. Layers produced by direct exfoliation were first identified using optical contrast in the microscope and imaged only few minutes after exfoliation in ambient conditions. A typical AFM image of a multilayer 2D-phosphane is shown in Figure 1a. The flake appears in AFM with a relative thickness of about 2.8 nm, as measured from the folded-back layer. Considering that the layer thickness of a 2D-phosphane monolayer is about 0.53 nm, the apparent height in AFM indicates the presence of roughly five layers[2]. Consistent with recent reports[13,17], degradation is noted by the homogeneous distribution of small bumps or defects at the surface of the layer in AFM images taken shortly after exfoliation. These bumps appear to grow and then collapse from the top of the layer rather than from the edges. The signs of degradation become obvious in AFM (Figure 1b) or under the optical microscope after few days. A drastic transformation is observed along with new droplets distributed in the vicinity of the initial position of the flakes. Obviously, degradation took place in those experimental conditions. Surprisingly, it was noted that degradation slowed down significantly when the samples was kept in the dark.

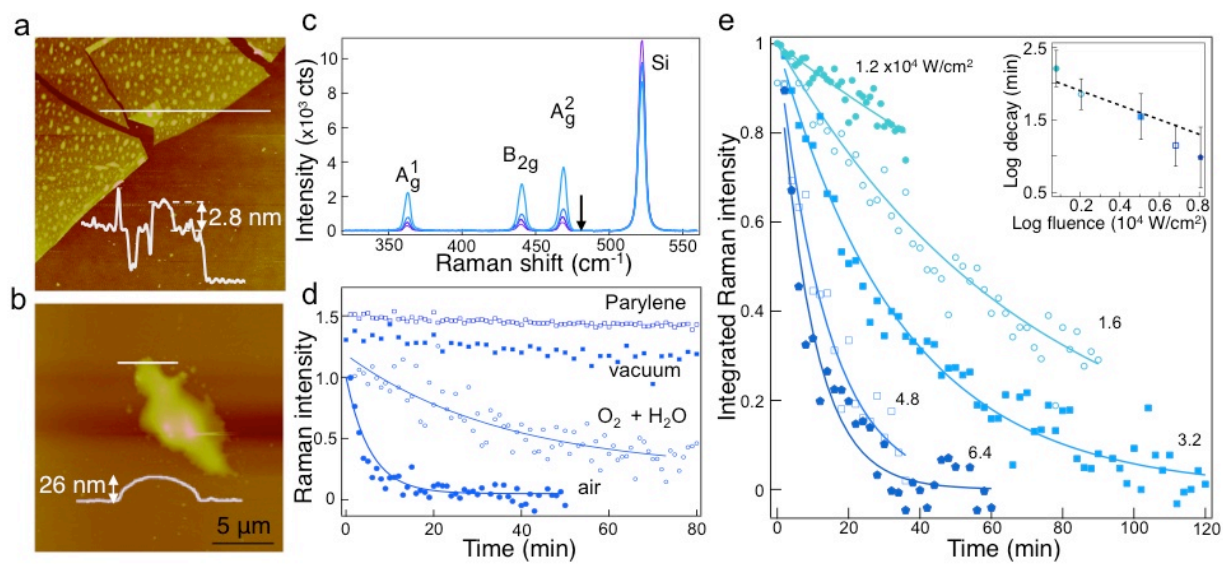

**Figure 1 | Photo-oxidation of thin layers of 2D-phosphane. a-b,** AFM images taken immediately after exfoliation on a SiO$_2$/Si substrate (**a**) and after a few days under ambient conditions (**b**). **c,** Raman spectra at λ = 532 nm measured in air 24, 48, 96 and 120 min after exfoliation. **d,** Time dependence of the integrated intensity of the $A_g^2$ Raman mode in different conditions: Air, vacuum, mixture of O$_2$ and H$_2$O, and in air but with a 20 nm capping layer of parylene C. Curves were vertically shifted for clarity. **e,** Time evolution of the intensity of $A_g^2$ mode for different laser fluences for a sample immersed in an aqueous solution in air. **Inset:** Intensity decay time as a function of laser fluence. The dashed line shows a slope of minus one.

Raman spectra of a 2D-phosphane multilayer are shown in Figures **1c** at 24, 48, 96 and 120 minutes after exfoliation and continuous exposure to a fluence of 2x10$^4$ Wcm$^{-2}$ at λ = 532 nm. Using established vibrational modes of bulk P(black), the observed Raman modes centered at 361 cm$^{-1}$, 438 cm$^{-1}$ and 466 cm$^{-1}$ are assigned to $A_g^1$, $B_{2g}$ and $A_g^2$, respectively[6,18-20]. As Raman scattering intensity is very sensitive to crystal quality, the quenching of the intensity within two hours is another indication of a fast degradation of the multilayer 2D-phosphane under light exposure.

An experiment was carried out in vacuum (< 5 x10$^{-6}$ Torr) in order to verify if the mere presence of visible light suffices to induce degradation. To minimize the influence of the hydrophilicity of the Si/SiO$_2$ substrate, a single crystal flake of multilayer 2D-phosphane was mechanically exfoliated and transferred onto a SiO$_2$/Si substrate coated with 20 nm of parylene C [21]. Raman spectra were recorded at room temperature as a function of time with and without gas exposure. No change in the Raman signal is observed for samples kept in vacuum (Figure 1d) or exposed separately to oxygen and water (at a partial pressure of 400 and 13 Torr, respectively, which is comparable to their partial pressure in air, Fig. S7). Experiments in vacuum using a higher flux of photons (up to 6 x10$^4$ Wcm$^{-2}$) also provided no evidence of degradation. The light-induced degradation of the multilayer 2D-phosphane is however instantly activated by introducing air or alternatively a mixture of pure oxygen and water vapor (Figure 1d). Moreover, the integrated Raman intensity as a function of time indicates that degradation in air is monoexponential, $\sim e^{-t/\tau}$, where $t$ and $\tau$ represent the time and a decay time, respectively. Note that different decay times are measured between experiments in air and in O$_2$/H$_2$O (5.5 and 36 min$^{-1}$, respectively) due to the different photon flux used (i.e. 1.7x10$^4$ Wcm$^{-2}$ and 1.8x10$^3$ Wcm$^{-2}$, respectively). Measurements of this kinetics in a buffer solution at pH between 5.8 and 7.8 indicated no dependency with pH. Finally, prolonged exposures to O$_2$, H$_2$O and then a mix O$_2$/H$_2$O were carried out in the dark and no degradation was noted, even after several hours (Fig. S7). All of these results reveal that three major environmental parameters are simultaneously required for the degradation of multilayer 2D-phosphane: water, oxygen and visible light.

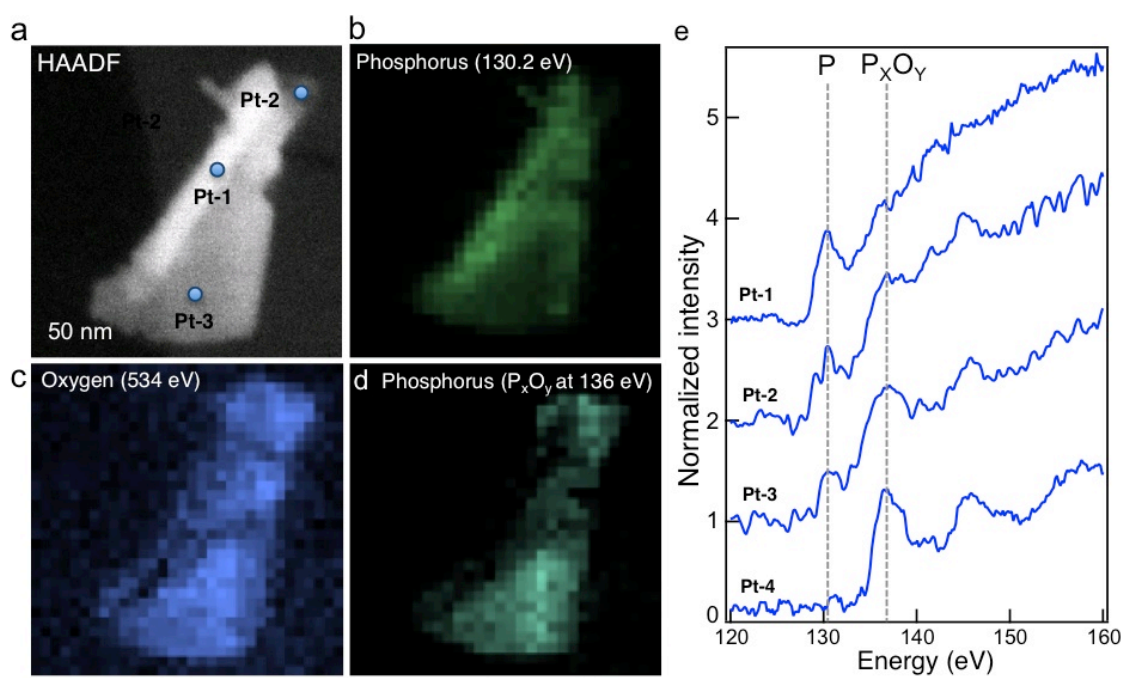

**Figure 2 | Chemical analysis by hyperspectral TEM-EELS spectroscopy of a multilayer 2D-phosphane. a** High Angle Annular Dark Field (HAADF) contrast image taken at 80 kV. **b-d**

EELS images extracted from the data cube at the energy of phosphorus L-2,3 edge (130.2 and 136 eV) and oxygen K-edge (534 eV). **e** EELS spectra corresponding to the selected areas in (a). Note that Pt 4 is acquired from another flake that is more oxidized, as described in Supplemental Information, Figure S10a.

Partially degraded samples were prepared for further TEM analysis of this chemistry using a combination of High Angle Annular Dark Field (TEM-HAADF) and hyperspectral electron energy loss spectroscopy (TEM-EELS) in imaging modes. HAADF contrast is directly linked to the number of atoms interacting with the electron beam, thus to the layer thickness, while EELS provides a local chemical analysis of the sample using signals at 130.2 eV, 136 eV and 534 eV that can be attributed to the presence of P, $P_xO_y$, and O, respectively (see Supplementary). In Figure 2a, the HAADF image of the multilayer 2D-phosphane flake indicates regions with for instance ~ 7 (Pt1), and ~ 2-3 (Pt2 and Pt3) 2D-phosphane layers (Fig. S11). The EELS spectra in these regions and the intensity map of the entire flake are displayed in Figure 2e and 2b-d, respectively. Pt1 region exhibits weak concentrations of oxygen and $P_xO_y$, species comparatively to the pristine P concentration. On the contrary, the presence of a significant amount of oxygen atoms preferentially localized in the thinnest regions of the flake is noted in Figure 2c. These regions are also characterized by the presence of pristine P and $P_xO_y$ species (Figure 2d), indicating partial oxidation. Recorded on another flake (Fig S10), the spectrum Pt4 in Figure 2e clearly demonstrates that the signature of pristine P at 130.2 eV can completely vanish at the cost of an increase of the $P_xO_y$ peak loss, indicating a complete degradation by oxidation reaction. These experimental results demonstrate that thin layers of 2D-phosphane are oxidized and the extent of the oxidation is thickness dependent and faster for ultra-thin layers.

From the results presented in Figures 1 and 2, a light-induced oxidative reaction mechanism can be sketched using the two main reaction steps involving the coverage, $\theta$, of a pristine and intrinsic 2D-phosphane:

$$\theta + h\nu \rightleftharpoons \theta^*, \qquad \text{Eq.1}$$

$$\theta^* + O_{2(aq)} \rightarrow O^{\cdot-}_{2(aq)} + \theta \rightarrow \theta_{ox}. \qquad \text{Eq.2}$$

In the first step, the optical excitation of the 2D-phosphane ground state produces a coverage of excited 2D-phosphane ($\theta^*$), whose steady-state population depends on the photon flux, the recombination rate and the absorption cross section (Eq. 1). $\theta^*$ is implicitly assumed here to involve precursors of free electron-hole pairs. In the second step (Eq. 2), the excited state of population $\theta^*$ undergoes a charge transfer reaction toward the aqueous oxygen molecules adsorbed at the surface of the layer, leading to p-doped 2D-phosphane. Finally, the charge transfer reaction with the oxygen-water redox couple generates reactive intermediate species (i.e. strong Brønsted bases), such as superoxide anions, that react spontaneously with the surface atoms of the 2D-phosphane layer and etch them in the process leading to oxidized species of coverage $\theta_{ox}$.

In analogy with a model explaining the ubiquitous air doping of carbon nanotubes and graphene layers deposited on an $SiO_2$ surfaces[22-24], the Marcus-Gerischer theory[25] (MGT) was used to model the reaction rate $\frac{d\theta}{dt}$, driven by the charge transfer from 2D-phosphane to oxygen/water, leading oxidation and etching (see supplementary for details and Fig. S12). This rate is given by:

$$\frac{d\theta}{dt} \propto -\theta J_{ph}[O_2] \cdot exp\left[-\frac{\left(\frac{|E_{g,n}|}{2}+E_i-E_{F,redox}^0+\lambda\right)^2}{4k_BT\lambda}\right], \qquad \text{Eq. 3}$$

where $\theta$ is the coverage of pristine 2D-phosphane, $J_{ph}$ is the laser fluence, $[O_2]$ is the oxygen concentration in water, $E_i$ is the intrinsic Fermi level, $E_{g,n}$ is the thickness dependent bandgap energy for a n-layer 2D-phosphane, $E_{F,redox}^0$ is the energy level of the oxygen acceptor state, and $\lambda$ is the renormalization energy of oxygen in water. This expression highlights the key role of the three main parameters on the reaction rate ($[O_2]$, $\theta$ and $J_{ph}$) and of the exponential dependency with the overlap between the 2D-phosphane density of states and the energy levels of the oxygen acceptor states in water.

As shown in Figure 1e, this model of the degradation process is tested using measurements on multilayer 2D-phosphane immersed in water (see Fig. S5) by recording the Raman signal as a function of time for a laser fluence between 1.2 x$10^4$ and 6.4 x$10^4$ Wcm$^{-2}$. As expected, the results show that the higher the photon flux, the faster the reaction. Indeed, on a log-log scale, the decay time as a function of fluence shows an approximate slope of -1 (see inset Figure 1e). Furthermore, the simple MGT explains the thickness dependence of the multilayer 2D-phosphane reactivity in terms of electronic confinement. When a thin sample is prepared, the bandgap shifts towards higher energies and closer to energy levels of the oxygen acceptor states, thereby strongly increasing the rate of charge transfer and hence the oxidation rate. This is especially true for the thinnest flakes, as observed in Figure 2. Other experiments (not shown) determined also that the reaction rate is independent of the pH of the solution because the reaction is driven by the $O_2/O_2^{\cdot-}$ as the P(black) is consumed before reaching equilibrium. MGT seems to fit quite well all of the observed behaviors and therefore proves that rigorous conditions are required in order to prepare thin layers of pristine (oxygen free) 2D-phosphane.

To prepare pristine layers, we devised an exfoliation procedure in the dark or inside a nitrogen-filled glove box in order to remove at least one of the components leading to degradation. The 2D-phosphane samples are transferred in those conditions onto a 300 nm SiO$_2$/Si substrate and the thinnest flakes are identified in the glove box by AFM and optical microscopy using a bandpass filter (λ >580 nm) to provide a more accurate value of the optical contrast. Finally, samples were directly transferred from the glove box to a vacuum chamber and then pumped down to $2.10^{-6}$ Torr for Raman experiments. Another strategy for experiemnts outside the glove box consists in coating the samples with a thin layer of a protecting insulator material. As an example in Figure 1d, 300 nm of parylene C is an excellent barrier against water adsorption and diffusion.

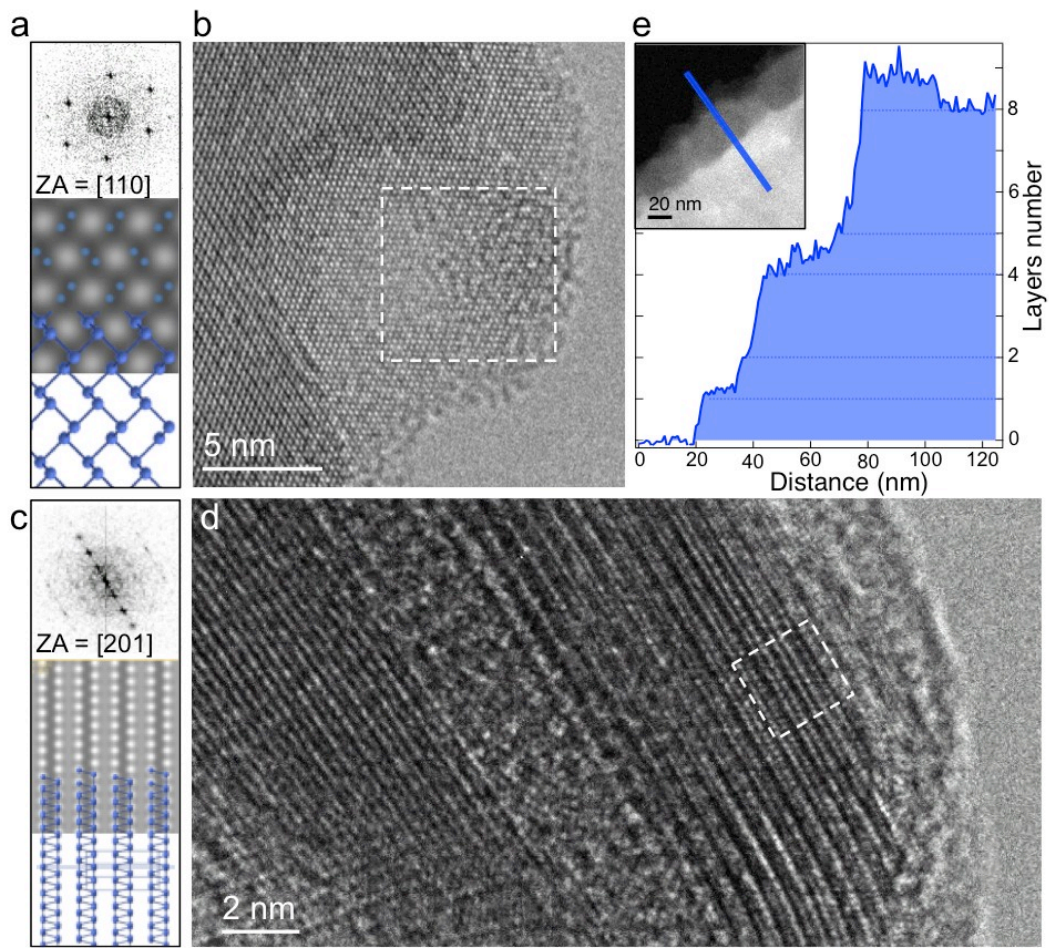

**Figure 3 | HRTEM images at 200kV of 2D-phosphane layers** viewed in the crystallographic projection along the *c*- axis ((110) zone axis corresponding to the edge-on view and slightly tilted (17°) top view) in (**a**) – (**b**) and along the *b*-axis ((201) zone axis) in (**c**) – (**d**): HRTEM images are shown in (**b**) and (**d**). Fast Fourier Transforms of the white dashed areas highlighted in images (**b**) and (**d**) are shown in (**a**) and (**c**) respectively together with corresponding simulated images of the structure models depicted in blue. **e** Evolution of the number of 2D-phosphane along the cross line in the inset extracted from the High Angle Annular Dark Field (HAADF) contrast image (inset) taken at the edge of a thin flake.

Pristine 2D-phosphane layers were prepared in dim light and quickly transferred on a TEM grid (see Supplementary Information) for structural analysis. Figure 3 presents TEM images of mono-, bi- and multilayer 2D-phosphane, Fast Fourier Transforms (FFT) of selected areas, and image simulations of the layered structure projected along different orientations. The exfoliation provided micrographs with top and side views. The puckered structure of 2D-phosphane is clearly revealed in Figure 3d, thanks to a fold in the layers that provides a view along their *b*-axis. The structure is confirmed by both the analysis of the FFT pattern of the selected area and the image simulation of the layered structure viewed along the (201) zone axis. The same can be said for the mono- and multilayer 2D-phosphane in Figure 3b and viewed along the *c*-axis ((110) zone axis). From the image simulations shown in Figure 3a, the stacking of the layers can be distinguished and monolayer 2D-phosphane identified. Furthermore, TEM-EELS core-loss analysis at the phosphorus K and L edges revealed that the sample essentially consists of pure phosphorus. Scanning TEM High Angle Annular Dark Field (STEM-HAADF, Figure 3e), which intensity is proportional to the number of phosphorus planes, was used to determine the thicknesses and the structural integrity of the 2D-phosphane samples. Overall, these TEM images demonstrate that pristine (with oxygen content of less than 1%) 2D-phosphane (composed of $n =1$ to 8 layers) can be readily produced when appropriate precautions are taken.

A large selection of samples produced in a glove box was characterized by AFM, optical contrast and Raman spectroscopy. As shown in Figure 4, the energy position and linewidth of Raman modes are consistent with those previously reported for bulk P(black)[18]. However, significant variations are observed as a function of thickness, from the bulk ($n = \infty$) down to monolayer 2D-phosphane ($n = 1$). The AFM image of this monolayer is shown in Fig. S6. The central frequencies of $A_g^1$ and $B_{2g}$ do not significantly shift, but a noticeable relatively symmetric broadening develops for $n \leq 5$. Down to the bilayer, widths increased by 1.5 cm$^{-1}$ with respect to the bulk. In contrast, the behavior of $A_g^2$ is both qualitatively and quantitatively different. As shown in Figure 5b, the central frequency of the $A_g^2$ shifts towards higher energies, reaches a maximum value for the bilayer (+2.6 cm$^{-1}$), and then red shifts such that the frequency of the monolayer is only +2 cm$^{-1}$ above that of the bulk. The full-width at half-maximum (FWHM) of $A_g^2$ follows a similar non-monotonic behavior. As shown in Figure 5a, the width increases continuously from the bulk value of 3.3 to 6.8 cm$^{-1}$ for the bilayer, and it then drops to 5.3 cm$^{-1}$ for the monolayer. These two non-monotonic dependencies strongly suggest the presence of more than one Raman allowed vibrational modes in the vicinity of $A_g^2$. This hypothesis is confirmed by the low-temperature bilayer spectra shown in the inset of Figure 4, where the presence of two modes can be unambiguously identified at 77 K.

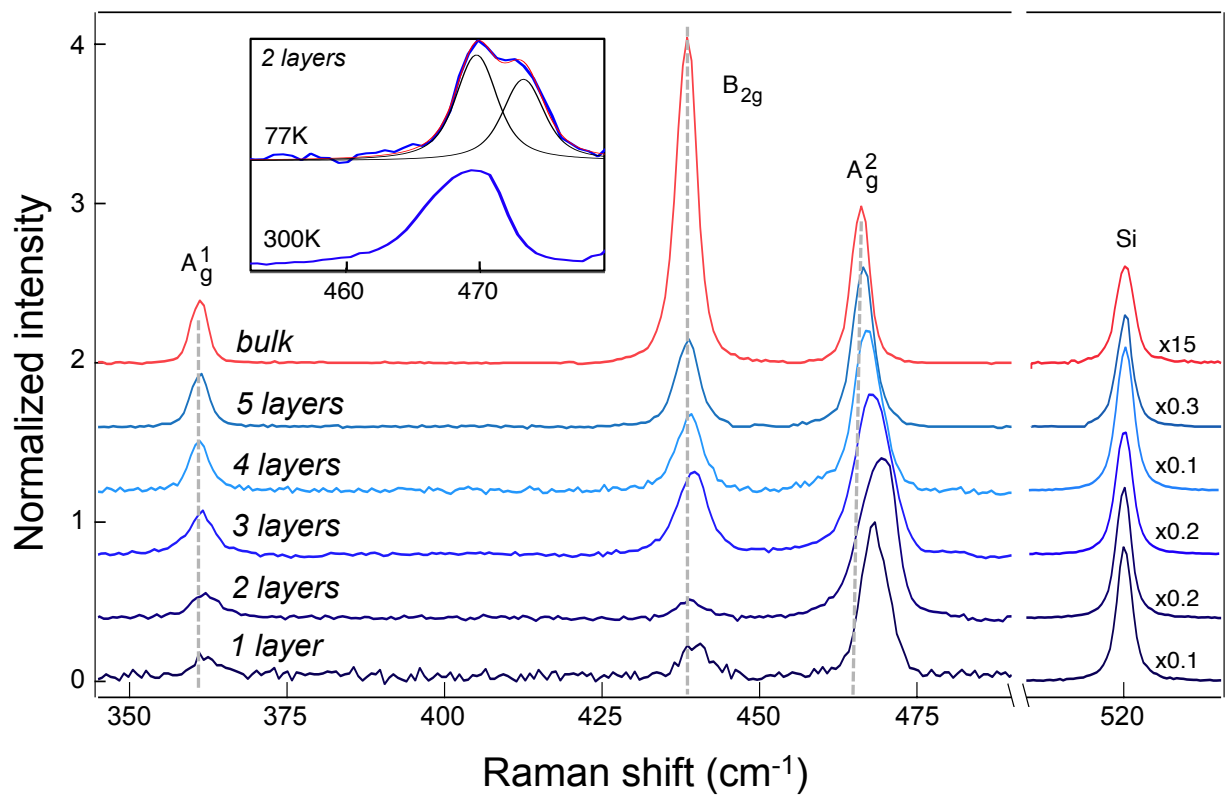

**Figure 4 | Raman spectroscopy of n-layer 2D-phosphane and bulk P(black).** Raman spectra at λ = 532 nm on layers of different thicknesses. Spectra are normalized to the $A_g^2$ intensity and vertically shifted for clarity (scaling factor on right is only for the Si peak region near 520 cm$^{-1}$). Inset : expanded view of the $A_g^2$ mode at 469 cm$^{-1}$ of the bilayer 2D-phosphane at 300K and 77K.

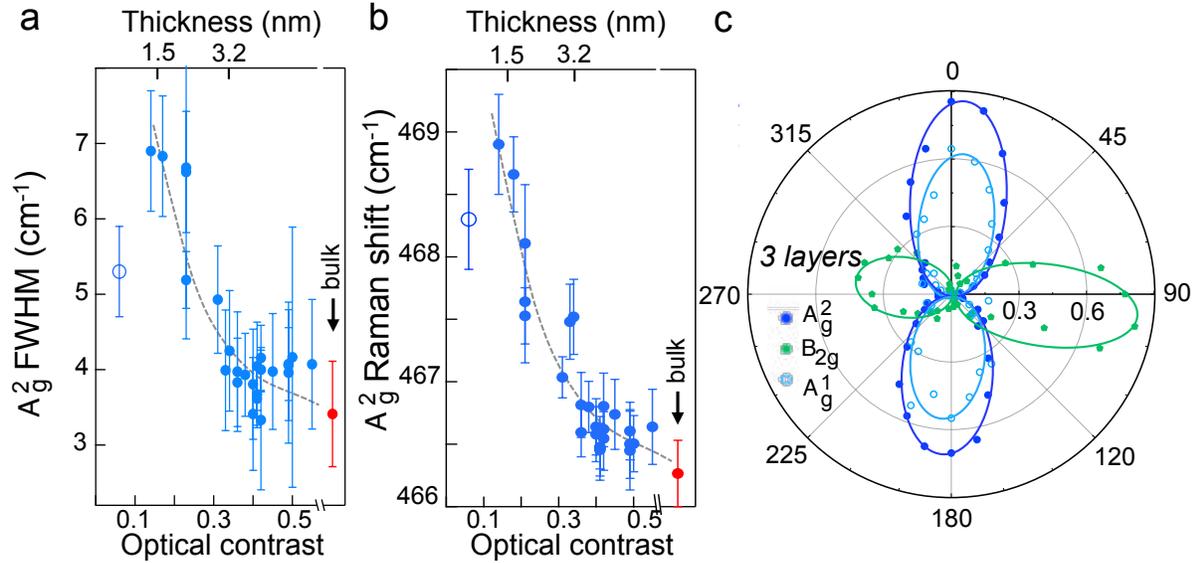

**Figure 5 | Raman characteristics of the $A_g^2$ mode of n-layer 2D-phosphane as a function of AFM thickness, optical contrast and laser polarization. a** and **b,** Evolution of the $A_g^2$ full-width-at-half-maximum (FWHM) (a) and of the spectral position (b) as a function of optical contrast. Open blue and solid red circles in (a) and (b) correspond to the monolayer 2D-phosphane and bulk P(black), respectively. **c,** Polarization-resolved integrated Raman intensities of $A_g^1$, $B_{2g}$ and $A_g^2$ from the trilayer (optical contrast of 0.22). The flake was excited with the polarization corresponding to the $A_g^2$ intensity maximum.

These Raman results can be understood on the basis of an analysis of the effect of the number of monolayers on the symmetry of multilayer 2D-phosphane. Bulk P(black) is orthorhombic (space group Cmce; no. 64) with the main crystallographic *b* axis normal to the layers. The unit cell is composed of 8 P atoms, disposed in two layers, each containing 4 P atoms that are covalently bonded through p-orbitals to three neighbors[26]. Consistent with the results in Figure 1c and Figure 4, the $D_{2h}$ factor group symmetry of bulk P(black) predicts three out of a total of six Raman active modes for an excitation along the crystal *b* axis and perpendicular to the two main (*a* and *c*) axes of the 2D-phosphane plane. The preparation of thin layers of 2D-phosphane breaks the translational invariance along *b* and reduces the number of symmetry operations. For samples with an odd or even number of monolayers, the symmetry is Pmna and Pbcm, respectively. Irrespective of the number of layers involved, the factor group always remains $D_{2h}$ and, conveniently, the same 8 one-dimensional representations can be used to label all vibrational modes. The broadening of the modes observed in Figure 4 can be associated with the Davydov splitting of monolayer vibrational modes and the appearance of new Raman-allowed modes[27,28]. Both effects result from the larger number of phosphorus atoms composing the primitive cell of bilayers, trilayers, and other few-monolayer samples compared to that of the monolayer 2D-phosphane and bulk P(black) primitive cells, which both contain only four atoms. Accordingly, multilayers are characterized by $6n$ Raman modes, half of which can be observed for scattered light along the *b* axis. Therefore, singly degenerate bulk modes and their corresponding monolayer modes are replaced, in few layers 2D-phosphane, by a number of near-degenerate modes corresponding to in- and out-of-phase stacking of monolayer vibrational motions[27]. Interestingly, some Raman modes of the multilayer are built from the combination of infrared (IR) modes in symmetrically opposed monolayers. Therefore, bulk and monolayer forbidden Raman modes can appear at energies close to that of bulk IR modes.

Davydov splitting requires at least $n = 3$, for which at least two Raman modes coexist in a narrow frequency range[27]. It therefore does not explain the broadening observed for the bilayer. For $A_g^1$ and $B_{2g}$, this broadening is weak and could be induced by various effects, from interactions with the substrate to inhomogeneous broadening on the scale of the diffraction-limited optical probe. For $A_g^2$ however, the degeneracy is clearly lifted, suggesting the presence of another mechanism. As described above, the presence of a second mode could be explained by combining two monolayer IR modes of $B_{2u}^1$ representation to produce a Raman-allowed $A_g^2$ mode. In the bulk, $B_{2u}^1$ is found at 469 cm$^{-1}$, which is very close to that of $A_g^2$ at 470 cm$^{-1}$. This symmetry analysis also predicts that all these near-generate modes should be absent in the monolayer because its primitive cell contains only 4 atoms. Indeed, we find that the width of $A_g^2$ of the monolayer is narrower than that of the bi- and trilayer, as was found for example for MoS$_2$[29].

The shift of $A_g^2$, a vibrational mode associated to atomic movements in the plane of the monolayer, bears resemblance to the hardening of $E_{2g}^1$ in MoS$_2$ with decreasing thickness[29]. Interestingly, the positions of $A_g^1$, assigned to atomic displacements perpendicular to the monolayer, and $B_{2g}$ do not significantly shift. Compared to transition metal dichalcogenides, interlayer interactions and their effects on vibrational energies appear to be qualitatively different. The frequency dependence of Raman modes is often used in 2D materials to reliably determine sample thickness with a monolayer precision. For 2D-phosphane, a similar procedure can be exploited, but only if Davydov-related effects highlighted above are properly taken into account. The monotonic dependence of intensity ratios on $n$ could be used to determine the number of monolayers involved, but the analysis must also carefully consider crystal anisotropy[30]. As highlighted in Figure 5c, Raman intensities are very sensitive to polarization of the excited and scattered beam. These polarization responses were obtained from the trilayer 2D-phosphane and respect the expected polarization selection rule.

In conclusion, a photo-induced oxidation process of 2D-phosphane samples was identified and shown to involve oxygen species in water present in air at the surface of the layer. The rate of the reaction was found to vary with the layer thickness, with the oxygen concentration and linearly with the light intensity. The oxidation rate is also predicted to depend exponentially on the square of the energy gap of the layer. A procedure was devised to exfoliate black phosphorus while preventing degradation in ambient conditions. The Raman signatures of mono-, bi- and multilayer 2D-phosphane were obtained and several thickness-related effects were identified and discussed using symmetry arguments. This work paves the way to a systematic investigation of the electrical and optical properties of pristine 2D-phosphane down to the monolayer.

**Materials and Methods:**

The black phosphorous (99.998%) was bought from Smart Elements (Vienna, Austria). Polydimethylsiloxane (PDMS) stamps were made using the PDMS solutions (Sylgard 184) bought from Dow Corning. Parylene C[19] was purchased from Cookson Electronics Specialty Coating Systems.

The 2D-phosphane layers were obtained by a modified scotch tape exfoliation technique[16] performed in a glove box under a clean nitrogen flow. First, the P(black) crystal is exfoliated onto a flat PDMS and then transferred onto a curved PDMS. The 2nd PDMS stamp covered with flakes is then rolled onto a 300 nm $SiO_2$/Si substrate (depicted in Supplemental, Figure S1).

For the degradation study, the exfoliation was done in a relatively dark room, and keeping the light very low, the flakes were characterized by optical contrast using a 580 nm low pass filter, followed by Raman spectrometry. For measurements under water, the pH was adjusted to 5.8 using phosphate buffer and the flakes were kept under water with a homemade liquid cell that is air tight for Raman measurements (See Supplemental Figure S5). For the optical contrast, AFM thickness and Raman correlation measurements, the degradation was avoided by placing an optical microscope and an atomic force microscope in a glove box and under continuous nitrogen flow. The exfoliation procedure of the n-layer 2D-phosphane and their AFM and optical characterizations were all performed in the same glove box. Sample transfers from the glove box to a cryostat for Raman scattering measurements were performed *in-situ* in order to avoid air exposure. To protect thin layers of 2D-phosphane during Raman characterization, the cryostat was continuously maintained at a pressure of $< 5.10^{-6}$ Torr.

**ACKNOWLEDGEMENTS**

The authors would like to acknowledge the assistance of Andreas Dietrich with early steps of sample preparation and characterization and discussions with Thomas Szkopek. The authors also acknowledge Patricia Moraille, from the Laboratoire de Caracterisation des Matériaux (LCM) at the Université de Montreal for her expertise on controlled atmosphere AFM measurements. This work was possible because of financial support from the Natural Sciences and Engineering Research Council of Canada (NSERC), the Fonds de Recherche du Québec-Nature et Technologie (FRQNT), and the Canada Research Chair (CRC) programmes. The research leading to these results has also received partial funding from the European Union Seventh Framework Program under grant agreement n°604391 Graphene Flagship.

Supplementary file for

# Exfoliating pristine black phosphorus down to the monolayer: photo-oxidation and quantum confinement

Alexandre Favron[1], Etienne Gaufrès[2], Fréderic Fossard[3], Pierre L. Lévesque[2], Anne-Laurence Phaneuf-L'Heureux[4], Nathalie. Y-W. Tang[2], Annick Loiseau[3] Richard Leonelli[1], Sébastien Francoeur[4] and Richard Martel[2]*

[1] Regroupement québécois sur les matériaux de pointe (RQMP) and Département de physique, Université de Montréal, Montréal PQ H3C 3J7, Canada

[2] RQMP and Département de chimie, Université de Montréal, Montréal PQ H3C 3J7, Canada

[3] Laboratoire d'Etude des Microstructures, UMR 104 CNRS-Onera, Châtillon, France

[4] RQMP and Département de génie physique, École Polytechnique de Montréal, Montréal PQ H3C 3A7, Canada

*Correspondence to: r.martel@umontreal.ca

## A. METHODS

### A1- Exfoliation of Black Phosphorous

Stratophosphane (2D-phosphane) mono-, bi- and multilayers were exfoliated from black phosphorus (P(black)) using a technique derived from the scotch-tape exfoliation technique[1]. The modified exfoliation process involved two polydimethylsiloxane (PDMS) stamps, as depicted in Figure S1. The first PDMS stamp was reticulated in a petri dish, while the second one was reticulated in a cylindrical tube of 1 cm diameter. The exfoliation was done in a dark room or in a glove box under nitrogen flux.

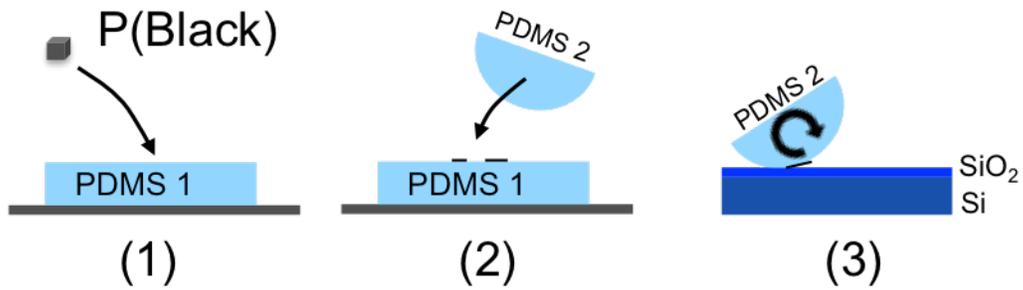

**Figure S1 | Three step exfoliation technique of P(black).** Step 1: Exfoliation done on the flat PDMS-1. Step 2: The flakes were reported on semi-spherical PDMS-2 stamp. Step 3: the stamp was rolled on the substrate (300 nm SiO$_2$ on a Si wafer) with an estimated speed of 0.1 cm/s.

### A2- Optical contrast measurements of 2D-phosphane layers

The flakes were localized and characterized under an optical microscope (Olympus BX51) placed in a glove box using red light (glass colored optical filter 580 nm) to limit the photo-oxidation of the flakes. The optical contrast, $C_{red}$, was measured preferentially at the center of the image and using the following formula:

$$C_{red} = \frac{R_{sub} - R_{flake}}{R_{sub}}, \qquad (S1)$$

where $R_{sub}$ and $R_{flake}$ are the reflection signals on the substrate and flake respectively. Some examples of 2D-phosphane layers thus produced are presented in Figure S2.

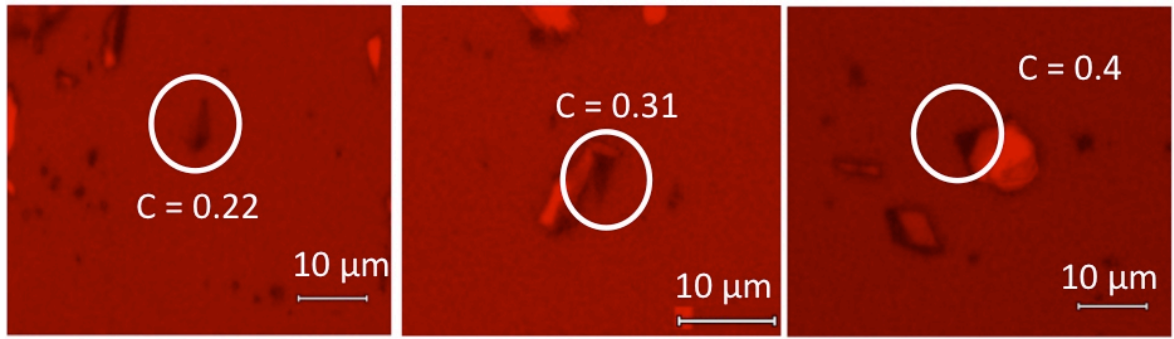

**Figure S2 | Examples of optical micrograph of the localization of flakes of 2D-phosphane multilayers using different optical contrasts.**

## A3- Measurements of the layer thickness using atomic force microscopy (AFM)

The substrate used is a 300 nm $SiO_2$ grown on a $n^{++}$ Si wafer that was previously patterned with metallic alignment marks (crosses done by photolithography). Short time before exfoliation, the substrate was annealed at 300°C for 3 hours and then quickly placed in a glove box equipped with an optical microscope fitted with a glass colored optical filter 580 nm and an AFM instrument (Auto Probe CP, ThermoMicroscopes). The exfoliation procedure as well as optical and AFM characterizations of the exfoliated P(black) flakes were all done in the same glove box and under a continuous flow of nitrogen (i.e. without contact to the atmosphere). The thicknesses of the flakes was measured using intermittent imaging mode at 85% damping and using silicon cantilever probes for tapping mode (ACTA from AppNano, Si probes with Al coating, tip radius < 10 nm, spring constant 25-75 N/m). The experimental procedure was done in the glove box to avoid degradation and to establish a correlation between the optical contrast and the average thickness of different flakes. The results of many different 2D-phosphane layers are presented in Figure S3.

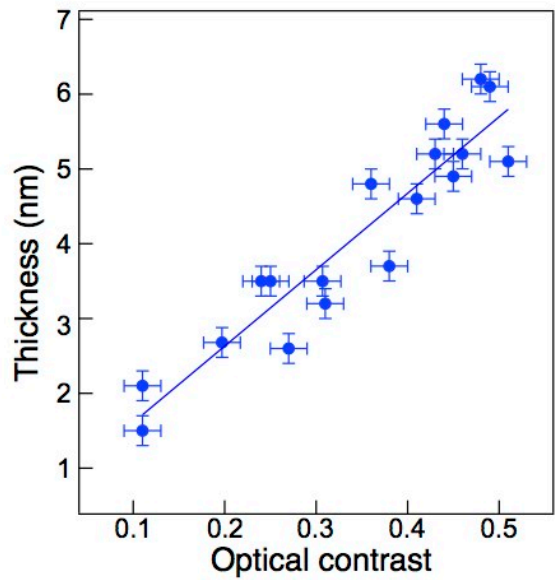

**Figure | S3 Evolution of the optical contrast of flakes as a function of the thickness**. The thickness of many 2D-phosphane thin layers with their different optical contrast values as measured by AFM and optical microscopy. The flakes are exfoliated in a glove box on 300 nm $SiO_2$/Si substrates.

## A4 - Photo-oxidation monitoring

A simple way to follow the progression of the photo-oxidation on the flakes was to compare a sequence of optical microscopic images at different exposure time to illumination in air. Several bumps appear after several minutes of exposition as shown in Fig. S4. AFM measurements in similar conditions also revealed the presence of bumps even before they can be seen in the optical microscope (See for instance Figure 1a in the main text)

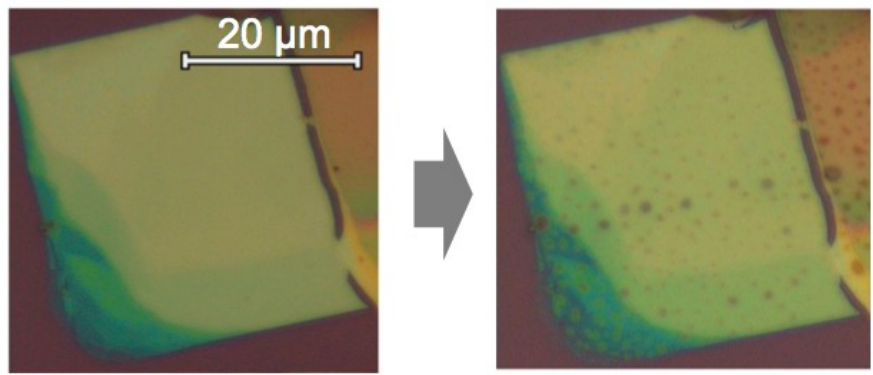

**Figure S4 | Optical microscopy images of a 2D-phosphane multilayer on a SiO$_2$/Si wafer taken just after exfoliation (left) and a few hours in air with light exposure (right).**

To quantify the photo-oxidation, we followed the integrated intensity of the Raman modes of P(black) modes, which is directly related to the coverage of 2D-phosphane multilayers in their pristine state. Hence, the oxidation of the surface will cause the Raman modes to decrease in intensity. The Raman spectrometer used is a custom built instrument with an excitation laser line at 532 nm (Centenia laser), a 50X objective and a nitrogen cooled charged-coupled device camera (JY Symphony). The spectral region probed was between 200-550 cm$^{-1}$ with a precision of ± 0.2 cm$^{-1}$.

**Raman measurements on multilayer 2D-phosphane flakes for degradation experiments:**

The sample stage was equipped with a cryostat (Janis ST500) coupled to a gas manifold and a pump. The vacuum pressure inside the cryostat chamber was below 2 x 10$^{-5}$ Torr. The gases introduced in the chamber were oxygen (leak of ~0.5 Torr) and water vapor at partial pressures of 475 Torr and 13 Torr, respectively. The deionized water used in this experiment was previously deoxygenated using several freeze thaw cycles during pumping.

**Raman measurements on samples immersed in water:**

As shown in the Figure S5, the samples were placed in a buffer solution maintained at pH of 5.8 using phosphoric acid and sodium hydroxide. To minimize the laser path length through the solution, a cover glass slide was placed close to the sample with a PDMS donut on top. The slide was isolated from the solution using a PDMS donut as shown in Figure S5.

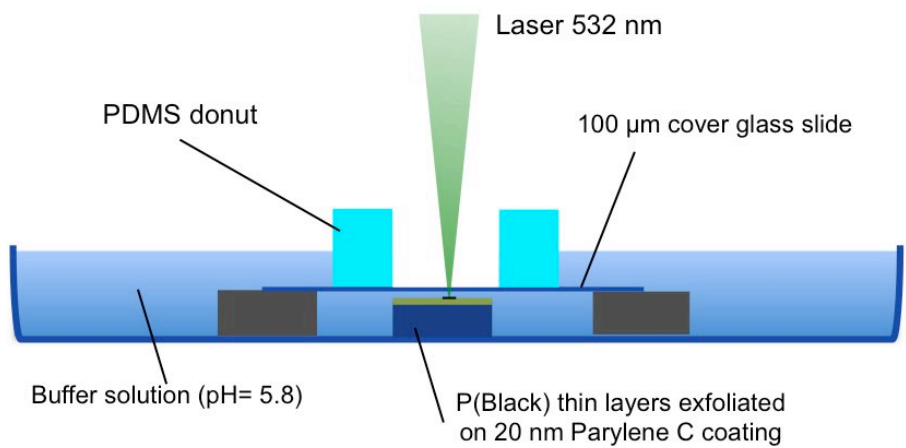

**Figure S5 | Scheme of the experimental setup for the Raman experiments of the 2D-phosphane multilayers in deionized water.**

## A5 – Raman measurement of the monolayer 2D-phosphane

The figure S6a presents the AFM image of a monolayer 2D-phosphane. The small dimension of the flake prevented measurement of optical contrast and the presence of a multilayer 2D-phosphane in contact with the monolayer imposed stringent conditions for the Raman measurements. The Figure S6b shows the position of the laser beam spot on the sample (532 nm, NA 0.55). It was placed carefully on the upper corner of the monolayer in order to avoid contamination of the signal by the multilayers. The very small part of the monolayer under the beam spot implied a long acquisition time (20 min) at a fluence of 100 $\mu W.\mu m^{-2}$.

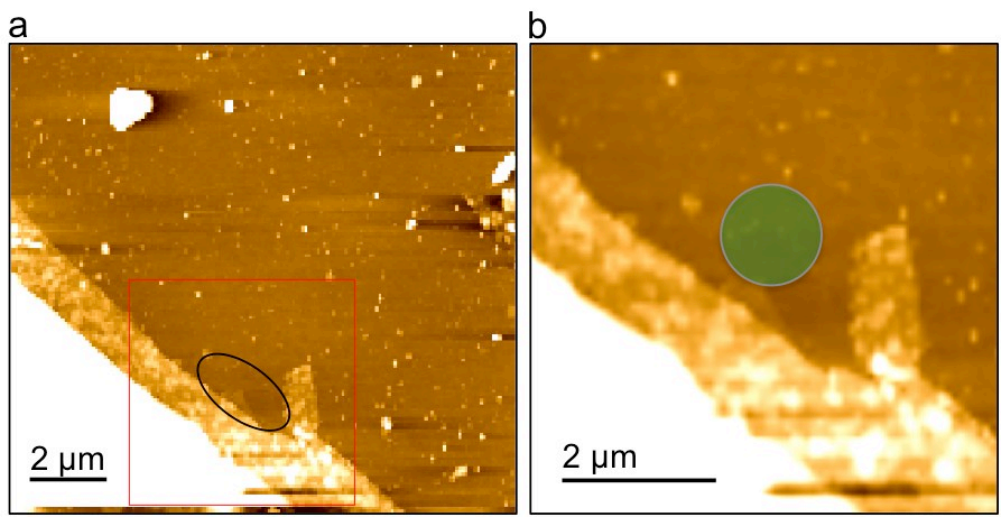

**Figure S6 | Experiments on the monolayer 2D-phosphane. a:** AFM image of a monolayer 2D-phosphane. **b:** Zoom of (a) and representation of the beam spot position on the monolayer 2D-phosphane for the Raman measurement in Figure 4, main text.

## B. ADDITIONAL EXPERIMENTAL RESULTS
### B1 – Raman measurements

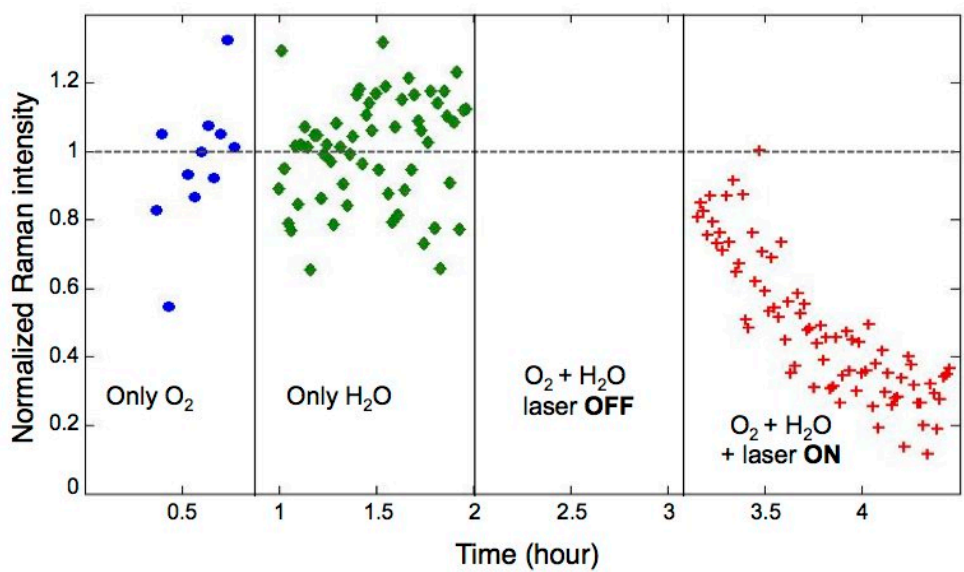

**Figure S7 | Evolution of the $A_g^2$ integrated Raman intensity of multilayer 2D-phosphane under different gas exposures.** The laser illumination is at a wavelength of 532 nm light source of 200 µW.µm$^{-2}$ intensity.

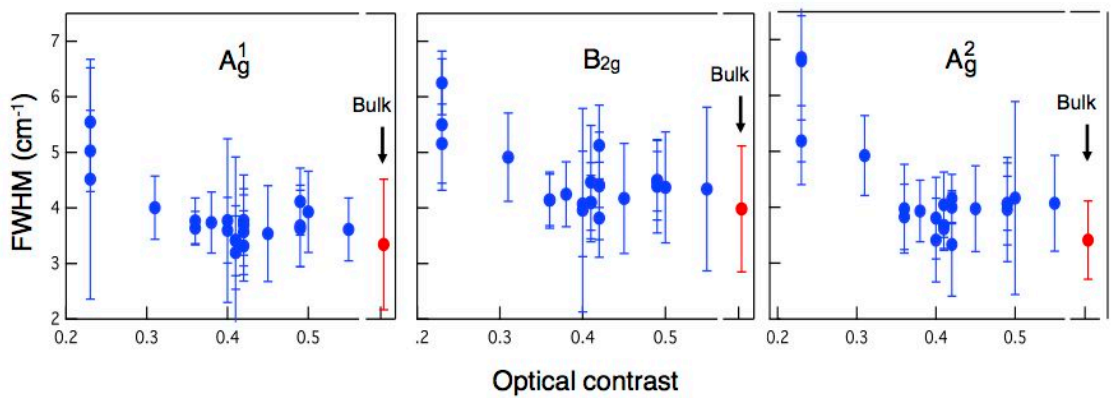

**Figure S8 | Evolution of the full width at half maximum (FWHM) of the $A_g^1$, $B_{2g}$ and $A_g^2$ Raman modes of 2D-phosphane layers of different thicknesses as a function of optical contrast.**

**B2– Transmission electron microscopy (TEM) imaging and spectroscopies**

**i) Microscope characteristics**

TEM and high-angle annular dark-field (HAADF) images were recorded with a Libra 200 MC Zeiss operating at 200 kV and at 80 kV. The microscope is equipped with a monochromatized Schottky FEG source delivering an energy resolution down to 150 meV and a Gatan Ultrascan camera. Point resolution is 2.3 Å and information limit is 1.2 Å at 200kV. The TEM can operate in scanning mode (STEM) with a spatial resolution under 1 nm and is equipped with Bright Field (BF) and High Angular Dark Field (HADF) detectors. Finally the machine is equipped with an in-column energy filter making possible energy filtered TEM, EELS measurements in Scanning TEM mode and acquisition of spectrum imaging datacubes (or hyperspectral EELS).

**ii) Exfoliation of P(black) on the TEM grid**

P(black) exfoliation was done using PDMS stamps in a low-light "dark" room. After three exfoliation steps, the grid was delicately placed in contact with the PDMS stamp to transfer thin layers on the carbon thin film of the grid for TEM analysis.

**iii) Simulations of High Resolution TEM (HRTEM) images**

Experimental HRTEM images of the 2D-phosphane layers were simulated using the JEMS software with the following TEM parameters:

|  | ZA [110] | ZA [201] |
|---|---|---|
| α (mrad) | 0.1 | 0.1 |
| Δf (nm) | Scherzer - 67±5 | Scherzer - 500±5 |
| HV (kV) | 200 | 200 |
| $C_S$ (mm) | 1.2 | 1.2 |
| $C_C$ (mm) | 1.2 | 1.2 |
| ΔE (eV) | <0.7 | <0.7 |

**iv) EELS measurements**

Figure S9 displays a core loss spectrum recorded at 80 kV on an almost pure sample used for TEM imaging in Figure 2. Assignments of the peaks are based on Refs (2) and (3). Phosphorous states are identified by inspecting the P-L2,3 edge. The main sharp P-L2,3 edge at 130.2 eV is the main signature of the P(black) structure. This peak is followed by a very weak peak at 136 eV assigned to an oxidized state corresponding to the $P_2O_5$ stoichiometry. Correlated with the peak at 136 eV, the spectrum displays a very weak peak at 534 eV corresponding to the K-O edge.

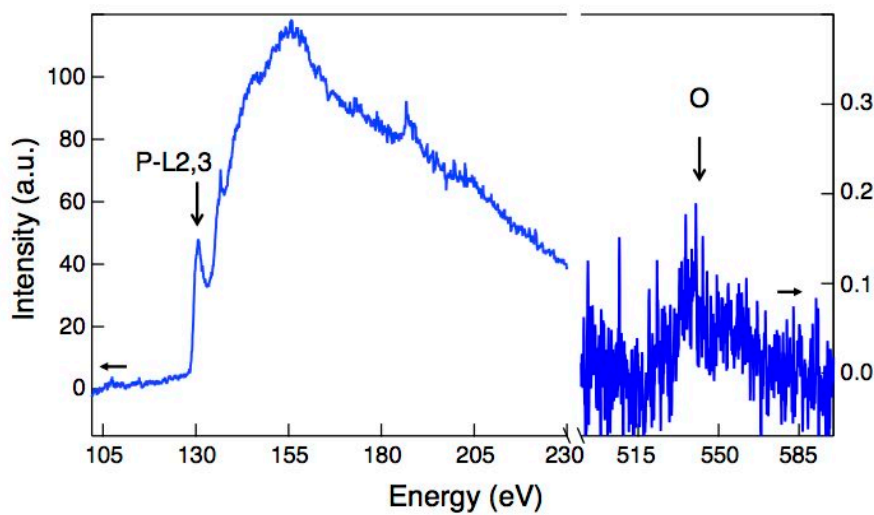

**Figure S9 | Core loss spectrum recorded at 80 kV and an energy resolution of 0.8 eV on a multilayer 2D-phosphane sample at P-L2,3 and O – K edges.**

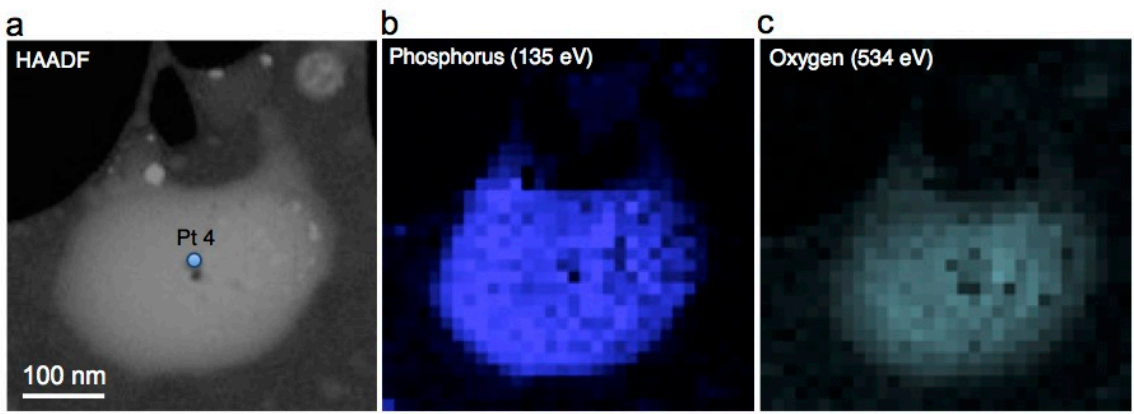

**Figure S10 | HAADF image (a) and core loss images at 136 eV (b) and 534 eV (c), recorded at 80 kV with an energy resolution of 0.8 eV. The Pt 4 mark in (a) corresponds to the location of the EELS spectrum displayed in Figure 2e of the main text.**

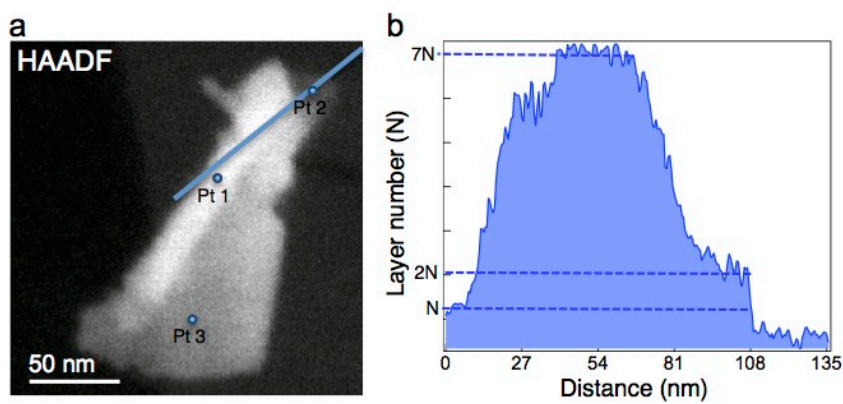

**Figure S11 | Evolution of the number of 2D-phosphane layers along the cross-section extracted from the High Angle Annular Dark Field (HAADF) contrast image (a) and used for the experiment in Figure 2 of the main text.**

**C. MARCUS-GERISCHER THEORY OF PHOTO-OXIDATION**

We have developed a simple model based on Marcus-Gerischer theory (MGT) in order to better explain the photo-induced oxidation process in 2D-phosphane layers and its dependency with layer thickness.

**C1 - Charge transfer doping**

Charge transfer between the 2D-phosphane layers and the environment is here assumed to take place through partial reaction steps of the well-known oxygen-water redox couple that is naturally present in air, in accordance with the following reaction:

$$O_{2(aq)} + 4e^- + 4H^+ \leftrightarrows 2H_2O.$$

It was previously established that this oxygen-water redox equilibrium reaction, which is pH dependent, can displace the Fermi level of a semiconductor, such as graphene, towards the Nerstian potential of the aqueous (aq) solution of oxygen through direct charge transfer (CT).[4] This reaction is driven by the potential difference between the two systems. In the case of P(black), the driving force is significant because the Nerstian potential of the redox reaction at pH 6 is -5.3 V (i.e. -5.3 eV potential energy relative to the vacuum level) whereas the workfunction of intrinsic P(back) is about 4.2 eV[5]. In the case of graphene, which workfunction is about 4.5 eV, CT reaction leads to p-doping because equilibrium is only reached when the Fermi level equals that of the Nerstian potential of the solution – i.e. at ~-5.3 eV – because there is an infinite reservoir of oxygen in air (i.e. oxygen concentration is constant due to Henry's law). In the case of P(black), this charge transfer reaction appears however shortcut at an intermediate step by the local oxidation of P atoms (see main text), which reaction leads to etching rather than doping.

**C2 - Rate limiting step**

For completion, the oxygen/water redox reaction above involves at least 4 intermediate steps (i.e. a total of 4 electrons transferred) and each step presents different kinetics and intermediate species. The first is an electron transfer from P(black) to the solvated oxygen acceptor state according to:

$$O_{2(aq)} + e^- \leftrightarrows O_{2(aq)}^{\cdot -}.$$

Because this reaction does not required neither $H^+$ nor $OH^-$, it is pH independent. Also, its reverse direction is not really favored since $O_{2(aq)}^{·-}$ species are reactive intermediates against P(black). In the MGT framework, this CT step is expected to be very slow because aqueous $O_2$ acceptor state is located at about -3.1 eV relative to vacuum[6], which is well above that of the P(black) band edges (See Figure S12). As a result of poor overlap, this CT step appears as the rate limiting step for the oxidation of P(black) in air. Two experimental facts described in the main text support this hypothesis: 1) CT rate as determined by Raman is very slow in the dark; 2) the photo-oxidation reaction is found to be independent of the pH of the solution, at least in the range between 5.8 and 7.8.

## C3 - Reaction kinetics based on MGT

### i) General expression and considerations about energy levels

To access the kinetics of the photo-induced CT reaction, we use the MGT describing the interaction of the solvent (water) with $O_{2(aq)}$ and $O_{2(aq)}^{·-}$. Because of fast water rearrangements due the presence of the charge, the level fluctuates in energy over a range of ± 1 eV,[7] which is expressed in the MGT using a total distribution (TD) function of the state that follows a Gaussian shape (See Figure S12). This CT rate is given by:

$$D_{O_2}(E) = [O_2] \cdot b_o \, exp\left[\frac{-\left(E - E_{F,redox}^0 + \lambda\right)^2}{4kT\lambda}\right], \qquad (S2)$$

where $[O_2]$ is the concentration of oxygen in water, $E$ is the potential energy (in eV), $E_{F,redox}^0$ is the electrochemical potential energy (or Fermi level of the solution), $\lambda$ is the reorganization energy, and $b_o$ is a normalizing parameter such that $\int D_{O_2}(E)dE = [O_2]$. In principle, the CT kinetics may be related to the overlap between the density of states of the n-layer 2D-phosphane and the solution, but the levels in the solution are not equivalent, in a conventional sense, to that a band. That is, an electron cannot be transferred to aqueous oxygen through a simple physical process, such as band-to-band tunneling, because molecular rearrangements are taking place during the course of the reaction[4]. It remains however possible, energetically, to transfer an electron from P(black) or n-layers 2D-phosphane with excess electrons to the oxygen state because the energy levels are close.

Because $E^0_{F,redox}$=-4.1 eV for $O_{2(aq)}/O^{·-}_{2(aq)}$ and workfunction of P(back) is about 4.2 eV (bandgap, $E_g$=0.35 eV), it appears clear in Figure S12 that intrinsic (undoped) P(black) should be roughly stable in normal condition, which is consistent with experimental observations of a relatively good stability of the crystal in air, as reported by Bridgeman more than 100 years ago[8].

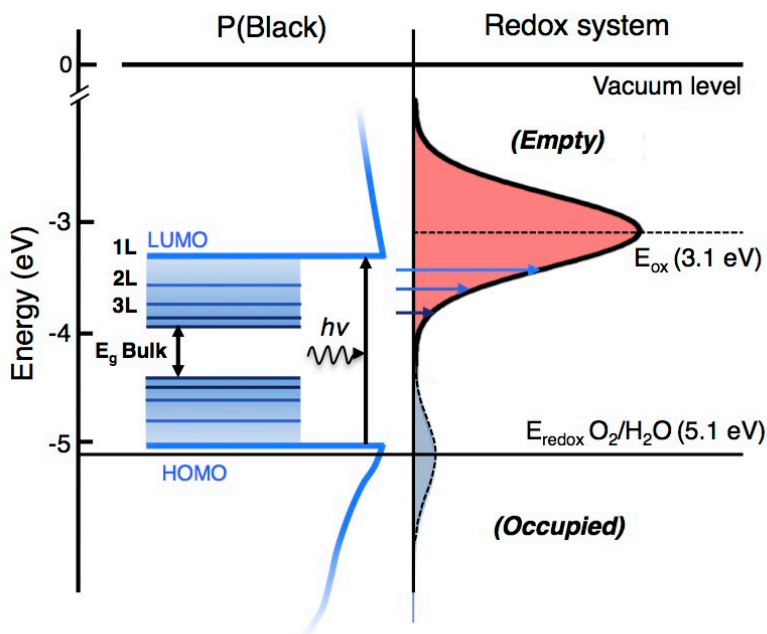

**Figure | S12 Sketch of the densities of states (DOS) of aqueous oxygen acceptor and of n-layers 2D-phosphane, as predicted by the Marcus-Gerisher theory**. **On the left**: DOS of mono-, bi-, tri- 2D-phosphane layers and bulk P(Black) extracted from Ref. 5 and 10. **On the right**: the energy diagram of the DOS of aqueous oxygen acceptor from Ref. (4) with the enlargement of the DOS due to solvent reorganization. The figure also illustrates by the blue arrows the proposed CT reaction induced by photo-excitation of n-layer 2D-phosphane.

### ii) Detail on the kinetics of charge transfer to bulk P(black)

The kinetic for the oxidation of the P(black) can be described by the following set of reactions. The overall reaction is:

$$\theta + O_{2(aq)} + h\nu \rightarrow \theta_{ox}\,. \quad \text{(S3)}$$

S3 can be subdivided into the following mechanism:

$$\theta + h\nu \xrightarrow{k_1} \theta^*, \qquad (S4)$$

$$\theta^* \xrightarrow{k_2} \theta + \Gamma_x, \qquad (S5)$$

$$\theta^* + O_{2(aq)} \xrightarrow{k_3} O_{2(aq)}^- + \theta + h^+, \qquad (S6)$$

$$O_{2(aq)}^- + \theta + h^+ \xrightarrow{k_4} \theta_{ox}. \qquad (S7)$$

In these equations, $\theta$ is the coverage of pristine P(black); $\theta^*$ is the coverage of excited P(black) by light – note that $\theta^*$ is also linked to the carrier concentration of photo-generated electrons, $\delta\eta$, at steady state because of the formation of electron-hole pairs; $\theta_{ox}$ is the coverage of oxidized P(black) in the form of $PO_2$; $h\nu$ is a photon count; $\Gamma_x$ regroups all decay channels leading to the relaxation of P(black) excited state to its ground state (phonon generation, fluorescence, etc.); $k_1, k_2, k_3, k_4$ are the rate constants.

Reaction S4 describes the direct photoexcitation of P(black). In this equation, we consider that electron-hole pairs are produced from the excitation and their relaxation towards the CB edge is very fast. Thus, carriers are generated and characterized by the energy of the band edges. Reaction S5 represents all the decay channels, radiative or not, of P(black) excited states. Driven by the CT to aqueous $O_2$, CT reaction S6 is responsible for the formation of the aqueous oxygen radial anion at the surface of P(black). This reaction is slow and assumed to be the limiting step with a rate constant, $k_3$. As detailed below, $k_3$ can be readily derived by the MGT. The last reaction S7 is a fast oxidation of P(black) by the oxygen anion oxidative specie or related intermediates, leading to etching.

Note that additional reactions are required in order to complete the reduction of $O_2$ towards $H_2O$ (i.e. a total of 4 electrons are involved in the complete water/oxygen redox reaction). However, our experiments indicate a high reactivity of $O_{2(aq)}^{\cdot-}$ intermediate species towards P(black) and thus any further electron transfer steps are neglected. To be noted, the reaction S6 and S7 are p-doping and de-doping processes of P(black), respectively. Thus, these steps overall lead to no doping of P(black). Furthermore, $O_{2(aq)}^{\cdot-}$ species can attack any pristine sites on the P(black) surface because the positive charges in reaction S6 is delocalized in the P(black).

To simplify the reaction kinetics, we assume that S4 and S5 are characterized by extremely fast kinetics compared to the CT processes in S6 and S7. Thus, these reactions can be considered as static, while solving S4 and S5 for the photo-stationary equilibrium situation, which sets the carrier population available for CT, as:

$$\frac{d\theta^*}{dt} = k_1\theta \cdot h\nu - k_2 \cdot \theta^* = 0 . \qquad (S8)$$

This yields:

$$\theta^* = \frac{k_1}{k_2}\theta \cdot h\nu . \qquad (S9)$$

From this equation, an excess of electron in the conduction band at steady state, $\delta\eta$, can be obtained for CT, corresponding to a coverage $\theta^*$. This excess is therefore directly proportional to the photon flux and surface coverage $\theta$. The concentration of thermally generated carriers is neglected in the actual process, which is justified by the significant bandgap of the material (of more than 0.35 eV in n-layer 2D-phosphane). This is consistent with the good stability of our samples in ambient conditions when kept in the dark.

One can develop further S9 for a thin layer of P(black) to get: $\delta\eta = g_{opt}\tau_\eta = (J_{ph}A\sigma_{abs}\rho_A\, n\, \theta)\, \tau_\eta$, where $g_{opt}$ is the optical generation, corresponding to $J_{ph}A\sigma_{abs}\rho_A\, n\, \theta$, and $\tau_\eta = \frac{1}{k_2}$ is the effective carrier lifetime (i.e. electron lifetime in the CB). In this expression, $J_{ph}$ is the laser fluence, A the unit area, $\sigma_{abs}$ is the absorption cross-section, $\rho_A$ is the phosphorus surface density, $n$ is the number of layer.

In the presence of the aqueous oxygen condensed at the surface of P(black), the electron transfers from the CB edge to the oxygen acceptor state of the solution proceeds according to reaction S6. This process generates aqueous superoxide anions, $O^{\cdot-}_{2(aq)}$, that react spontaneously with the surface atoms of the 2D-phosphane layer and etch them in the process. We can then assume that the oxidation reaction will be mainly limited by the S6 reaction step.

In the MGT framework, the electron transfer rate between the conduction band of P(black) and the acceptor states of the solution, described by S6, is explicitly given by:

$$\frac{dO^{\cdot-}_{2(aq)}}{dt} = \int_{-\infty}^{\infty} b_i\theta^*\bigl(1 - f(E)\bigr)D_{O_2}(E)dE. \qquad (S10)$$

Where $D_{O_2}$ is given by S2, $b_i$ regroups several prefactors related to the geometry of the molecular system and $f(E)$ is the Fermi-Dirac distribution. Assuming now that all electrons are distributed at the minimum of the CB of a n-layer 2D-phosphane, $E_{c,n}$, by a Dirac delta function, we obtain after integrating:

$$\frac{dO_{2(aq)}^{\cdot -}}{dt} = b_i \theta^* D_{O_2}(E_{c,n}), \quad (S11)$$

or more explicitly using $\delta\eta = \theta^*$ and equation S2:

$$\frac{dO_{2(aq)}^{\cdot -}}{dt} = b_i b_o [O_2]\, \delta\eta\, exp\left[-\frac{(E_{c,n}-E_{F,redox}^0+\lambda)^2}{4kT\lambda}\right]. \quad (S12)$$

Because the rate equation for reaction S6 is: $\frac{dO_{2(aq)}^{\cdot -}}{dt} = k_3 \theta^* [O_2]$, the reaction constant $k_3$ therefore become:

$$k_3 = b_i b_o exp\left[-\frac{(E_{c,n}-E_{F,redox}^0+\lambda)^2}{4k_B T\lambda}\right]. \quad (S13)$$

Last, we can also write Equation S12 by taking into account that $\frac{dO_{2(aq)}^{\cdot -}}{dt} = \frac{d\theta_{ox}}{dt} = -\frac{d\theta}{dt}$. By renaming the variable $b \equiv b_i b_o$, the oxidation rate can be formulated by the following kinetic equation:

$$\frac{d\theta}{dt} = -b\delta\eta\,[O_2] exp\left[-\frac{(E_{c,n}-E_{F,redox}^0+\lambda)^2}{4k_B T\lambda}\right]. \quad (S14)$$

The above expression can also be rewritten in terms of the energy gap of a n-layer 2D-phosphane, $E_{g,n}$, assuming an intrinsic Fermi level, $E_i$, *i.e.* in the middle of the gap. With the definitions above, we can develop more explicitly the expression for the oxidation rate of the P(black):

$$\frac{d\theta}{dt} = -b(A\sigma_{abs}\rho_A \tau_n n\, k_0)\, \theta J_{ph}\, [O_2] \cdot exp\left[-\frac{\left(\frac{|E_{g,n}|}{2}+E_i-E_{F,redox}^0+\lambda\right)^2}{4k_B T\lambda}\right]. \quad (S15)$$

As discussed in the main text, the above expression describes well the observed behavior for the photo-oxidation of the P(black) and of n-layer 2D-phosphane. That is, the Raman intensity is proportional to the coverage $\theta$, which then decays exponentially with time. The oxidation rate is also directly proportional to the photon flux, the $O_2$ concentration (or partial pressure) and more strikingly, as evidenced by TEM-EELS results, the rate is fastest for the thinnest n-layer 2D-phosphane.

**C5 - On the consequences of electronic confinement on CT kinetics**

The situation of n-layer 2D-phosphane is different than for bulk P(black) because electronic confinement in n-layer 2D-phosphane leads to an enlargement of the band gap energy[9,10]. According to MGT, larger bandgaps should indeed improve the energy level alignment between the 2D-phosphane conduction band and the solvated oxygen acceptor state (see illustration in Figure S12). Hence, MGT predicts faster kinetics for electron transfer reaction toward the solution acceptor states (i.e. $O_{2(aq)}$) with decreasing thickness of the layer. In other words, the rate of transfer from the conduction band increases exponentially with the square of the bandgap energy, as indicated in Equation S15. Note that this is true up to the maximum level of the acceptor states, which is roughly -3.1eV. This situation is illustrated in Figure S12 by the relative length of the arrows in blue.

For intrinsic 2D-phosphane layers kept in the dark, the electron population in the conduction band is rather limited at room temperature and mainly given by the intrinsic carrier concentration. This population decreases with increasing bandgap. This explains the relatively good stability of thin 2D-phosphane layers in the dark in ambient condition. As discussed in the main text, n-layer 2D-phosphane exposure to light will generate a significant carrier population that is photo-generated by the process. The electrons and holes accumulate at the conduction and valence band edges and become available for transfer to the aqueous oxygen acceptor states.